\documentclass[a4paper,12pt]{article}
\usepackage{amssymb}
\usepackage{amsmath}
\usepackage{amstext}
\usepackage{amsfonts}
\usepackage{epsfig}
\usepackage{dcolumn}
\usepackage{rotating}
\usepackage{color}
\usepackage{comment}
\usepackage{hyperref}

\def\XXint#1#2#3{{\setbox0=\hbox{$#1{#2#3}{\int}$ }
\vcenter{\hbox{$#2#3$ }}\kern-.56\wd0}}

\newcommand*\xbar[1]{%
  \hbox{%
    \vbox{%
      \hrule height 0.5pt 
      \kern0.5ex
      \hbox{%
        \kern-0.1em
        \ensuremath{#1}%
        \kern-0.1em
      }%
    }%
  }%
}

\definecolor{rosso}{cmyk}{0,1,1,0.4}
\definecolor{rossos}{cmyk}{0,1,1,0.55}
\definecolor{rossoc}{cmyk}{0,1,1,0.2}
\definecolor{blu}{cmyk}{1,1,0,0.3}
\definecolor{blus}{cmyk}{1,1,0,0.6}
\definecolor{bluc}{cmyk}{1,1,0,0.1}
\definecolor{verde}{cmyk}{0.92,0,0.59,0.25}
\definecolor{verdec}{cmyk}{0.92,0,0.59,0.15}
\definecolor{verdes}{cmyk}{0.92,0,0.59,0.7}

\newcommand{\ba}{\begin{eqnarray}}
\newcommand{\ea}{\end{eqnarray}}
\newcommand{\be}{\begin{equation}}
\newcommand{\ee}{\end{equation}}
\newcommand{\bi}{\begin{itemize}}
\newcommand{\ei}{\end{itemize}}

\newcommand{\ga}{\gamma}

\newcommand{\da}{\delta}
\newcommand{\la}{\lambda}

\newcommand{\sa}{\sigma}
\newcommand{\en}{\epsilon}

\newcommand{\Ga}{\Gamma}

\newcommand{\cF}{{\cal F}}

\newcommand{\cM}{{\cal M}}

\newcommand{\ra}{\rightarrow}
\newcommand{\Ra}{\Rightarrow}

\newcommand{\LF}{\left(}
\newcommand{\RF}{\right)}
\newcommand{\LT}{\left[}
\newcommand{\RT}{\right]}

\newcommand{\kb}{\bar{k}}
\newcommand{\pb}{\bar{p}}

\newcommand{\4}{\frac{1}{4}}

\newcommand{\mt}{\mathtt}

\begin{document}

\title{Unitarity in Infinite Derivative Theories}
\author{Spyridon Talaganis \\ \\
 {\it Physics Department,} \\
{\it Lancaster University, Lancaster,} \\
{\it LA$1$ $4$YB, United Kingdom.}\\
\begin{footnotesize}\textit{E-mail}:  s.talaganis@lancaster.ac.uk \end{footnotesize}}

\date{}

\maketitle

\begin{abstract}
In this paper, we consider an infinite derivative scalar field action with infinite derivative kinetic and interaction terms. We establish that the theory is unitary if the correlation functions are formulated in Euclidean space and then analytically continued in their external momenta to Minkowski space.
\end{abstract}

\tableofcontents

\section{Introduction}
\numberwithin{equation}{section}

In many quantum gravitational theories, \textit{i.e.}, string theory~\cite{Polchinski:1998rr}, loop quantum gravity~\cite{Ashtekar,Nicolai:2005mc}, causal set~\cite{Henson:2006kf},  one can see that \emph{non-locality} is present in many of these theories of quantum gravity; for example, strings and branes are nonlocal by nature~\cite{Eliezer:1989cr}. In string field theory~\cite{Witten:1985cc,Siegel:1988yz}, nonlocality also plays an important role ($p$-adic strings~\cite{Freund:1987kt}, zeta strings~\cite{Dragovich:2007wb} and strings quantized on a random lattice~\cite{Douglas:1989ve}). Therefore, non-locality seems to play an important role in nature.

Inspired from string field theory~\cite{Witten:1985cc,Siegel:1988yz}, a covariant
\emph{non-local} gravitational theory free from ghosts and tachyons around constant curvature backgrounds was derived in Refs.~\cite{Biswas:2011ar,Biswas:2013kla}. The form of the action $S$ is given by
\begin{align}\label{aglar}
S & = S _{EH}+S_Q \,, \\
S_{EH} & = \frac{1}{2} \int d^{4} x \, \sqrt{-g} M_{P}^{2}  R \,, \\
S_{Q} &= \frac{1}{2} \int d^{4} x \, \sqrt{-g} \LF R \cF_{1}(\bar{\Box})R+R_{\mu \nu} \cF_{2}(\bar{\Box})R^{\mu \nu}+R_{\mu \nu \la \sa} \cF_{3}(\bar{\Box})R^{\mu \nu \la \sa} \RF \,,
\end{align}
where $\bar{\Box}\equiv \Box/M^2$ and $M$ is the mass scale at which the nonlocal modifications become important. The $\cF_{i}$'s are infinite-derivative functions of $\bar{\Box}$ and follow a specific constraint $2\cF_1(\bar{\Box})+\cF_2(\bar{\Box})+2\cF_3(\bar{\Box})=0$ around a Minkowski background so that the action is~\textit{ghost-free} and corresponds to a massless graviton~\cite{Biswas:2011ar,Biswas:2013kla}. In particular, the graviton propagator is modulated by the exponential of an \textit{entire function} $a(-k^2) = e^{k^2/M^2}$, see~\cite{Biswas:2005qr},
\be \label{spyros}
\Pi(-k^2) = \frac{1}{k^2a(-k^2)}\LF {\cal P}^2 - \frac{1}{2} {\cal P}_s ^0  \RF=\frac{1}{a(-k^2)} \Pi_{GR} \,.
\ee
Note that the exponential of an entire function does not give rise to poles. For an exponential entire function, the propagator becomes exponentially suppressed in the UV while the vertex factors are exponentially enhanced.
In~\cite{Talaganis:2017dwo}, the Slavnov identities for the infinite derivative gravitational theory whose action is given by~\eqref{aglar} were established. Therefore, the UV divergences of Feynman diagrams can be eliminated up to $2$-loop order~\cite{Talaganis:2014ida} and the theory is renormalisable~\cite{Talaganis:2017tnr}. Higher loops can also be made finite by the use of dressed vertices and dressed propagators. Meanwhile, in the IR, we recover the physical graviton propagator of GR. In addition, this asymptotically free theory addresses the classical singularities present in GR~\cite{Biswas:2011ar,Biswas:2014tua,Biswas:2013cha}. This is in clear contrast with GR and other finite-order higher-derivative theories of gravity. 

In~\cite{Talaganis:2016ovm}, the UV behaviour of scattering diagrams within the context of an infinite-derivative scalar toy model was investigated and it was found that the external momentum dependence of the scattering diagrams is convergent for large external momenta. 
In~\cite{Talaganis:2017dqy}, it was found, for an infinite derivative scalar toy model, that, as the number of particles increases, the corresponding effective mass scale associated with the scattering amplitude decreases.
In~\cite{ham}, the Hamiltonian for an infinite derivative gravitational theory was written down and the number of degrees of freedom in various cases was evaluated. Various aspects of infinite derivative theories were looked into in~\cite{Conroy:2015nva,Teimouri:2016ulk,Talaganis:2017evj}.

Unitarity plays a very important role in determining the viability of a theory.
Unitarity in non-local theories has been studied in~\cite{Biswas:2014yia,Tomboulis:1997gg,Modesto:2011kw,Addazi:2015ppa,Tomboulis:2015gfa,Alebastrov:1973vw,Anselmi:2017lia}.
In~\cite{Carone:2016eyp}, it was shown that a unitary, non-local theory can be formulated if the correlation functions are formulated in Euclidean space and then analytically continued in their external momenta to Minkowski space.
Following this treatment, we would like to verify the unitarity of an infinite derivative scalar field theory with infinite derivative kinetic and interaction terms.

The outline of the paper is as follows. In section~\ref{sec:feynman}, the Feynman rules for our scalar field theory are written down.
In section~\ref{sec:unitar}, the unitarity of the theory is established.

\section{Feynman Rules for Infinite Derivative Scalar Toy Model}
\numberwithin{equation}{section}
\label{sec:feynman}


Let us consider an infinite derivative scalar toy model whose action is given by~\cite{Talaganis:2014ida,Biswas:2014tua,Talaganis:2016ovm,Talaganis:2017tnr}
\be \label{dacho}
S_{\mt{scalar}} = S _ {\mt{free}} + S _ {\mt{int}}\,,
\ee
where
\be
S_{\mt{free}} = \frac{1}{2}\int d^4 x \, \LF  \phi \Box a(\bar{\Box}) \phi\RF
\ee
and
\be
S_{\mt{int}} = \frac{1}{M_P} \int d ^ 4 x \, \LF \frac{1}{4} \phi \partial _ {\mu} \phi \partial ^ {\mu} \phi + \frac{1}{4} \phi \Box \phi a(\bar{\Box}) \phi - \frac{1}{4} \phi \partial _ {\mu} \phi a(\bar{\Box})  \partial ^ {\mu} \phi \RF\,,
\ee
where we have that 
\be \label{sicko}
a (\bar{\Box}) = e^{\ga \bar{\Box}^{m}} \equiv e^{\ga \LF \frac{\Box}{M^2} \RF^{m}}\,.
\ee 
$\ga >0$ and $m$ is an even positive integer.

The Feynman rules for our action, which is given by Eq.~\eqref{dacho}, can be derived rather straightforwardly. The propagator in momentum space is then given in Euclidean space by
\be
\Pi_{E} (k ^ 2)= \frac{- i}{k^2 e ^ {\ga \kb ^ {2m}}}\,,
\ee
where barred $4$-momentum vectors from now on will denote the  momentum divided by the mass scale $M$. The vertex factor for three incoming momenta $k_{1},~k_{2},~k_{3}$ satisfying the conservation law:
\be
k _ {1} + k _ {2} + k _ {3} = 0\,,
\label{conservation}
\ee
is given by
\be
\label{eq:V}
\frac{1}{M_{P}}V (k _ {1}, k _ {2}, k _ {3}) = \frac{i}{M_P} C(k_1,k_2,k_3) \LT 1 -  e ^ {\ga \kb _ {1} ^ {2m}} -  e ^ {\ga \kb _ {2} ^ {2m}} - e ^ {\kb _ {3} ^ {\ga 2m}}\RT\,,
\ee
where
\be
C (k_1,k_2,k_3)= \frac{1}{4} \LF k _ {1} ^ {2} + k _ {2} ^ {2} + k _ {3} ^ {2} \RF\,.
\ee
Let us briefly explain how we obtain the vertex factor.
The first term originates from the term, $ \4 \phi \partial _ {\mu} \phi \partial ^ {\mu} \phi$, which using Eq.~\eqref{conservation} in the momentum space, reads
\ba
- \frac{i}{2} (k _ {1} \cdot k _ {2} + k _ {2} \cdot k _ {3} + k _ {3} \cdot k _ {1})
=  \frac{i}{4} \left(k _ {1} ^ {2} + k _ {2} ^ {2} + k _ {3} ^ {2} \right) \,.
\ea
The second term comes from the terms, $\frac{1}{4} \phi \Box \phi a(\Box) \phi$, and $- \frac{1}{4} \phi \partial _ {\mu} \phi a(\Box)  \partial ^ {\mu} \phi$. In the
momentum space, again using Eq.~\eqref{conservation}, we get
\be
\frac{i}{4} \left(k _ {3} \cdot k _ {1} + k _ {1} \cdot k _ {2} - k _ {3} ^ {2} - k _ {2} ^ {2} \right) e ^ {\ga \kb _ {1} ^ {2m}}  = - \frac{i}{4} \left(k _ {1} ^ {2} + k _ {2} ^ {2} + k _ {3} ^ {2} \right) e ^ {\ga \kb _ {1} ^ {2m}} \,.
\ee
 The third and the fourth terms in Eq.~\eqref{eq:V} arise in an identical fashion.

\section{Unitarity}
\numberwithin{equation}{section}
\label{sec:unitar}

Unitarity means that $S^{\dagger}S=1 \Ra i(T^{\dagger}-T)=T^{\dagger}T$ between the $S$-matrix and the $T$-matrix, where $S=1+iT$.

In~\cite{Carone:2016eyp}, it was shown, for a non-local theory, that, if correlation functions are
formulated in Euclidean space and analytically continued in the external momenta to Minkowski space, unitarity is not violated. That is, the $k^0$ integration is along the imaginary axis and the Euclidean external momenta are given by $p^{0}=ip_{E}^{0}$.

By using the aforementioned prescription and applying the optical theorem, we deduce that the theory is unitary (at least up to $1$-loop order). Even if we had chosen a different infinite derivative interaction term, for instance~\cite{Talaganis:2016ovm,Talaganis:2017dqy},
\be
S_{\mt{int}}=\la \int d^4 x \, \LF \phi \Box \phi a(\bar{\Box}) \phi \RF \,,
\ee
$a(\bar{\Box})$ being given by~\eqref{sicko} and $\la$ a coupling constant, the theory would still be unitary as whether unitarity holds or not is determined by the locations of the poles, which stay the same regardless of the interaction terms.

\section{Concluding Remarks}
\numberwithin{equation}{section}
\label{sec:concl}

The aim of this paper has been to show that infinite derivative theories
containing infinite derivative kinetic and interaction terms can be made unitary. If the correlation functions are formulated in Euclidean space and then analytically continued in their external momenta to Minkowski space, unitarity is preserved. For future work, it would be interesting to verify unitarity within the framework of an infinite derivative gravitational theory.

\section{Acknowledgements}

ST is supported by a scholarship from the Onassis Foundation.

\end{document}